\PassOptionsToPackage{table}{xcolor}
\documentclass[sigconf,nonacm]{acmart}
\usepackage{textcomp}
\usepackage{multirow}
\usepackage{makecell}
\usepackage{balance}
\usepackage{enumitem}
\setcopyright{none}
\renewcommand\footnotetextcopyrightpermission[1]{}

\newcommand{\subsub}[1]{\par\textbf{#1.}\ }

\title{Tailor: Accelerating Heterogeneous Multi-Scalar Multiplications in Zero-Knowledge Proof Systems}

\author{Gaurav Kuwar, Alhad Daftardar, Jianqiao Mo, and Brandon Reagen}
\affiliation{
    \institution{New York University Tandon School of Engineering}
    \city{Brooklyn}
    \state{NY}
    \country{USA}
}

\begin{document}

\begin{abstract}

Zero-knowledge proof generation is computationally expensive, with multi-scalar multiplication (MSM) often dominating prover runtime. Existing MSM accelerators primarily target long vectors of full-width scalars using Pippenger's algorithm. However, our characterization of six ZKP protocols reveals significantly heterogeneous MSMs: matrix MSMs that reuse points across rows, small-valued scalars, and sequences of short MSMs. Pippenger's bucket and window reductions are well amortized for long vector MSMs, but become disproportionately expensive for short MSMs and when repeated across matrix rows. Alternative MSM algorithms can reduce or eliminate these reductions, with precomputation-based methods further benefiting from point reuse. These observations expose a mismatch between the diversity of MSMs and architectures built around a fixed algorithm and configuration.

We present \textit{Tailor}\footnote{Our architecture is \textit{tailored} to each MSM workload.}, a programmable MSM architecture that adapts its algorithm, window size, and effective scalar width to each MSM at runtime. Tailor unifies multiple MSM algorithms on a shared accumulation datapath through configurable scalar decoding and accumulator partitioning. Its matrix dataflow reuses points across rows and interleaves independent reductions to improve utilization. Under a common area budget, Tailor achieves geometric-mean speedups of $3.1\times$ on dense matrix MSMs, $17.3\times$ on 16-bit matrix MSMs, and $1.25\times$ on dense vector MSMs over a matrix-adapted state-of-the-art accelerator. Integrated into a full-prover accelerator, Tailor achieves up to $13.16\times$ end-to-end speedup over a baseline system, with only the MSM unit replaced.

\end{abstract}

\maketitle

\section{Introduction}
Zero-Knowledge Proofs \cite{zkp} (ZKPs) allow a prover to convince a verifier that a computation was performed correctly without revealing private inputs. For example, a client can outsource computation to an untrusted service provider and verify the returned result is correct without repeating the computation. ZKPs enable applications such as verifiable Large-Language Model (LLM) inference \cite{zkllm, zkgpt}, general-purpose verifiable computing \cite{sok_zkvm_gap, jolt_paper}, blockchain scalability \cite{hawk}, and electronic voting.
However, their broader adoption is limited by the substantial computational cost of proof generation.

A core building block of ZKP protocols is Multi-Scalar Multiplication (MSM), which computes a dot product of scalar-point pairs on an elliptic curve. MSM accounts for 60--90\% of runtime in prior end-to-end ZKP accelerators \cite{cuzk, szkp, pipezk, zkspeed, zkphire, legozk, kuwar2026locus} and 23--89\% of CPU prover runtime across the protocols we study (Section~\ref{sec:msm_workload_char}). Consequently, MSM has attracted considerable ASIC \cite{szkp, pipezk, priormsm, zkspeed, gypsophila, myosotis, chipletmsm/252, legozk, reZK}, GPU \cite{cuzk, distmsm, elasticmsm, gzkp}, and FPGA \cite{cyclonemsm, bstmsm, hardcamlmsm, intelifzkp, msmac, pipemsm, flexmsm} acceleration research. Existing accelerators primarily optimize Pippenger's algorithm \cite{pippenger} for long vectors of full-width scalars, where bucket and window reductions are amortized over many scalar-point pairs.

Yet modern ZKP protocols generate highly heterogeneous MSMs, even within a single proof. Our characterization of six protocols (Section~\ref{sec:msm_workload_char}) reveals substantial variation in size, shape, and scalar characteristics. Multilinear polynomial commitment schemes such as Hyrax \cite{hyrax} and Dory \cite{dory_pcs_paper} produce matrix MSMs, comprising many independent short rows of scalars that share a common point vector. Other operations involve binary or small-valued scalars requiring far fewer bits than the curve's full scalar width. Furthermore, several protocols issue hundreds of short MSMs containing only a few points; in Spartan, these account for up to 57\% of prover runtime. For such computations, Pippenger's repeated bucket and window reductions can become disproportionately expensive. Alternative algorithms can reduce or avoid these reductions, either through precomputation and point reuse or through direct scalar multiplication.

The choice of algorithm and configuration also determines how hardware resources should be organized. Window size changes the balance between accumulator storage and reduction overhead, while precomputation trades point storage for fewer arithmetic operations. Matrix MSMs add another dimension, exposing independent rows that enable point reuse and parallel execution. Efficiently supporting these combinations requires an architecture that can adapt its resource allocation and execution dataflow.

We present \textit{Tailor}, a programmable, characteristic-aware MSM architecture that adapts its algorithm, window size, and effective scalar width to each computation. We observe that heterogeneous MSMs favor different algorithms and configurations, yet their varying execution patterns can be efficiently supported by a shared accumulation datapath through flexible resource allocation. Tailor realizes this through configurable scalar decoding and a dynamically partitioned accumulator pool, mapping independent accumulations across buckets, windows, and rows onto shared hardware. Its matrix-aware dataflow further reuses points across rows and interleaves independent reductions to improve pipeline utilization.

\begin{figure*}
    \centering
    \includegraphics[width=1\linewidth]{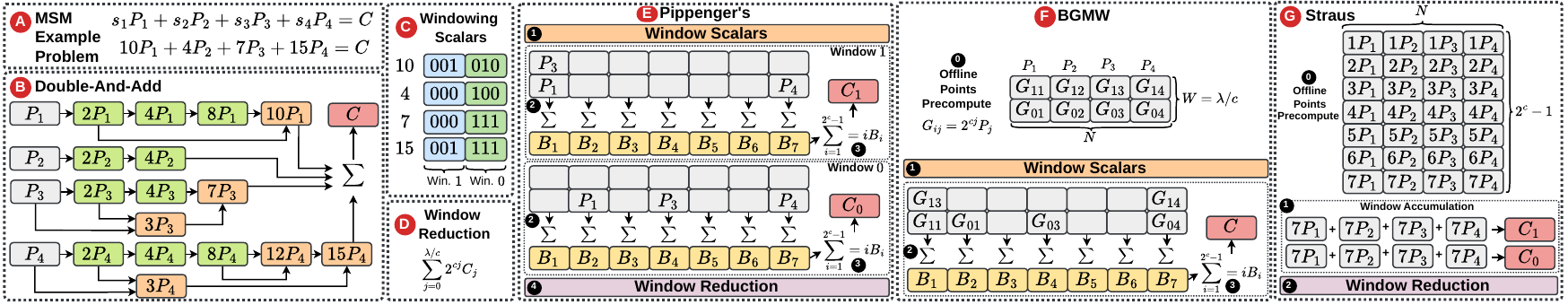}
    \vspace{-20pt}
    \caption{Computation of the example MSM problem (A) with each MSM algorithm. Double-and-Add (B) scans each scalar bit-by-bit, doubling per bit and adding when the bit is set. Windowing (C) splits scalars into $c$-bit digits recombined by Window Reduction, with (D) applying (B) per window. Pippenger's (E) bucket sorts each window, then reduces buckets and windows. BGMW (F) precomputes position-weighted points, needing only one bucket sort and reduction. Straus (G) precomputes digit multiples, reducing each window to lookup and accumulation.}
    \label{fig:algo_theory}
\vspace{-1.5em}
\end{figure*}

Our key contributions are:
\begin{itemize}[leftmargin=*]
    \item \textbf{MSM characterization.} We characterize MSMs across six ZKP protocols, identifying how size, shape, and scalar width affect the choice of algorithm and configuration.
    \item \textbf{Unified programmable architecture.} We develop a shared accumulation datapath with configurable scalar decoding and accumulator partitioning, together with a matrix-aware dataflow that exploits row- and window-level parallelism.
    \item \textbf{ASIC performance.} At a common area budget, Tailor achieves geometric-mean speedups of $3.1\times$ on dense matrix MSMs, $17.3\times$ on 16-bit matrix MSMs, and $1.25\times$ on dense vector MSMs over a matrix-adapted state-of-the-art accelerator. On MSMs from real ZKP protocols, Tailor achieves a $4.7\times$ geometric-mean speedup.
    \item \textbf{End-to-end acceleration.} Integrated into a full prover accelerator, Tailor achieves geometric-mean proving speedups of $10.2\times$ and $1.16\times$ on Spartan and HyperPlonk, respectively, over a baseline system with only the MSM unit replaced.
\end{itemize}

\section{Background}
\subsection{Zero-Knowledge Proofs}
Zero-Knowledge Proofs (ZKPs) \cite{zkp} allow a prover to convince a verifier that a statement is true without revealing anything beyond its truth. A prominent class of ZKPs is the zero-knowledge Succinct Non-interactive Argument of Knowledge (zkSNARK), which proves that a hidden witness satisfies algebraic constraints while providing compact proofs and fast verification \cite{zkspeed, zkphire, groth16}. Many modern zkSNARKs \cite{jolt_paper, hyperplonk, spartan_zkp} combine two components: a polynomial interactive oracle proof (PIOP), which reduces correctness to checking polynomial identities, and a polynomial commitment scheme (PCS), which allows the prover to commit to polynomials and prove their evaluations. Elliptic-curve-based PCSs \cite{kzg_pcs, dory_pcs_paper, hyrax} rely on computationally expensive operations, particularly multi-scalar multiplication (MSM), for commitments and openings, making MSM a major contributor to prover runtime.

\begin{figure}[t]
    \centering
    \includegraphics[width=1\linewidth]{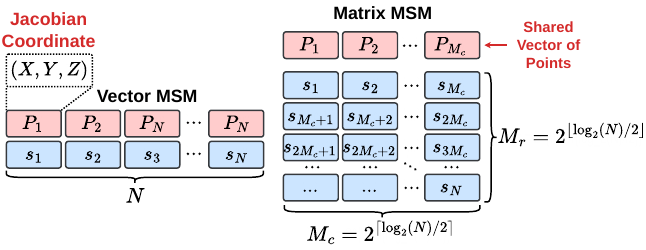}
    \vspace{-22pt}
    \caption{Vector and Matrix MSM shape for problem size $N$. Matrix MSMs share a smaller point vector across rows.}
    \label{fig:msm_shape}
\vspace{-1.25em}
\end{figure}
\begin{figure*}
    \centering
    \includegraphics[width=1\linewidth]{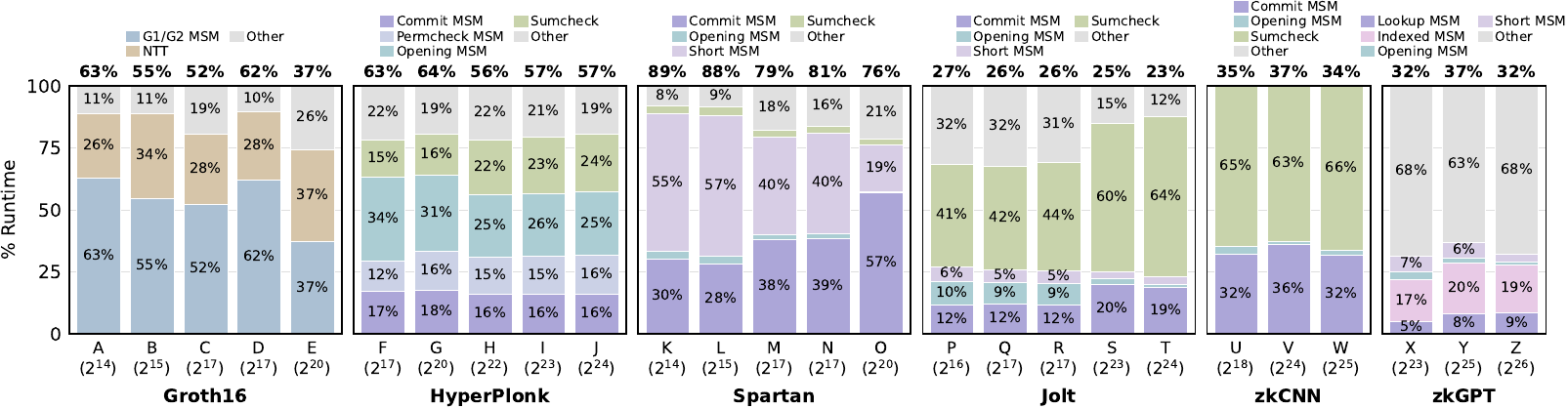}
    \vspace{-20pt}
    \caption{Prover runtime breakdown across ZKP protocols and workloads (labeled A--Z; as-per \autoref{tab:workload_letters}). Total MSM runtime percentage is annotated above each bar. All protocols run with 32 threads, except zkCNN (single-threaded).}
    \label{fig:combined_breakdown_single}
\vspace{-1.25em}
\end{figure*}

\subsection{Multi-Scalar Multiplication}\label{sec:msm}
Multi-scalar multiplication (MSM) computes a dot product of scalars and elliptic-curve points. Given $N$ scalars $(s_1, \dots, s_{N})$ and points $(P_1, \dots, P_{N})$, it evaluates:
$$
\sum_{i=1}^{N} s_i P_i
$$

MSM is computationally expensive because scalar multiplication is performed via \textit{repeated point additions}, each involving tens of modular multiplications. Scalars are typically around 256 bits wide, while elliptic-curve points are represented using two or three coordinates, each 254--381 bits wide.

\begin{table}[t]
\footnotesize
\centering
\caption{MSM Characteristics Taxonomy.}
\vspace{-7pt}
\label{tab:taxonomy}
\setlength{\tabcolsep}{2mm}
\begin{tabular}{|c|c|p{4.85cm}|}
\hline
\multicolumn{2}{|c|}{\textbf{Characteristic}} & \textbf{Description}   \\ \hline \hline
\multicolumn{2}{|c|}{\textbf{Size ($N$)}} & The number of scalars in the MSM. \\ \hline
\multirow{2}{*}{\textbf{Shape}} & Vector & A single MSM over a vector of $N$ scalars and points. \\ \cline{2-3}
 & Matrix & $N$ scalars typically arranged as $2^{\lfloor \log_2(N)/2 \rfloor}$ rows by $2^{\lceil \log_2(N)/2 \rceil}$ columns. Each row is an independent MSM sharing a single point vector. \\ \hline
\multicolumn{2}{|c|}{\textbf{Scalar Width ($\lambda$)}} & Bitwidth of the scalars (16b, 254b, 255b, etc.). \\ \hline
\multirow{3}{*}{\textbf{Distribution}} & Dense & Uniformly random scalars in $[2^{\lambda-1}, 2^{\lambda}-1]$. \\ \cline{2-3}
 & Small & Scalars within a narrow distribution centered at zero (2--70b), that may be negative. \\ \cline{2-3}
 & Binary & $0$ or $1$ scalars. \\ \hline
\multicolumn{2}{|c|}{\textbf{Sparsity}} & The fraction of scalars that are zero. \\ \hline
\end{tabular}
\vspace{-1.5em}
\end{table}

Although prior accelerators primarily focus on Pippenger's algorithm, several MSM algorithms exist (\autoref{fig:algo_theory}), each with different trade-offs. These algorithms differ in how they expose parallelism and exploit precomputation. In many ZKP protocols, the MSM base points are fixed at setup, so precomputation is a one-time offline cost, amortized over all proofs. \autoref{fig:algo_theory} illustrates the algorithms discussed below.

\subsub{Double-and-Add (D\&A)} This approach computes each scalar-point multiplication independently using the scalar's binary representation. For a $\lambda$-bit scalar, it performs successive point doublings and conditional point additions based on the scalar's bits. Point doubling uses a separate formula that typically requires fewer modular multiplications than point addition. Each scalar-point multiplication requires $O(\lambda)$ elliptic-curve operations, and the resulting points are accumulated to obtain the final MSM.

\subsub{Windowing Scalars} A common strategy partitions each $\lambda$-bit scalar into $W=\lceil\lambda/c\rceil$ windows of $c$ bits. We refer to a partition's position within the scalar as its \textit{window} and its value as a \textit{digit}. This transforms the MSM into $W$ independent per-window MSMs that can execute in parallel. Their results are combined through \textit{Window Reduction} using Horner's method \cite{horners}. Although typically associated with Pippenger's algorithm, windowing also applies to other MSM algorithms.

\subsub{Pippenger's \cite{pippenger}} Pippenger extends windowing with bucketing to reduce the number of point additions within each window. Points with the same nonzero digit $d$ are accumulated into a bucket $B_d$, requiring $2^c-1$ buckets per window. \textit{Bucket Reduction} then computes the weighted sum $\sum_d dB_d$ without adding each bucket $d$ times. This is performed independently for every window, followed by \textit{Window Reduction}. Both reductions are incurred for every MSM (or matrix row), with Bucket Reduction repeated across windows. Pippenger's reductions are well amortized for long MSMs, but can become disproportionately expensive for short MSMs or small scalars.

\begin{table}[t]
\small
\centering
\caption{Evaluated ZKP protocols and benchmarks, with problem sizes, elliptic curves, and PCS.}
\vspace{-7pt}
\label{tab:workload_letters}
\setlength{\tabcolsep}{1mm}
\renewcommand{\arraystretch}{1.1}
\resizebox{\columnwidth}{!}{%
\begin{tabular}{|c|c|l|c||c|c|l|c|}
\hline
\textbf{Protocol} & \textbf{Letter} & \textbf{Workload} & \textbf{Size} &
\textbf{Protocol} & \textbf{Letter} & \textbf{Workload} & \textbf{Size} \\ \hline \hline
\multirow{5}{*}{\makecell{\textbf{Groth16}\\BN254\\no PCS}} & A & AES & $2^{14}$ & \multirow{5}{*}{\makecell{\textbf{HyperPlonk}\\BLS12-381\\Multilinear\\KZG}} & F & Zcash & $2^{17}$ \\ \cline{2-4}\cline{6-8}
 & B & SHA2 & $2^{15}$ & & G & Auction & $2^{20}$ \\ \cline{2-4}\cline{6-8}
 & C & RSA & $2^{17}$ & & H & Zexe's Recursive Ckt & $2^{22}$ \\ \cline{2-4}\cline{6-8}
 & D & Merkle & $2^{17}$ & & I & Rollup of 10 Pvt Tx & $2^{23}$ \\ \cline{2-4}\cline{6-8}
 & E & Auction & $2^{20}$ & & J & Rollup of 25 Pvt Tx & $2^{24}$ \\ \hline
\multirow{5}{*}{\makecell{\textbf{Spartan}\\BLS12-381\\Hyrax}} & K & AES & $2^{14}$ & \multirow{5}{*}{\makecell{\textbf{Jolt}\\BN254\\Dory}} & P & BTree & $2^{16}$ \\ \cline{2-4}\cline{6-8}
 & L & SHA2 & $2^{15}$ & & Q & SHA2 & $2^{17}$ \\ \cline{2-4}\cline{6-8}
 & M & RSA & $2^{17}$ & & R & SHA3 & $2^{17}$ \\ \cline{2-4}\cline{6-8}
 & N & Merkle & $2^{17}$ & & S & Fibonacci & $2^{23}$ \\ \cline{2-4}\cline{6-8}
 & O & Auction & $2^{20}$ & & T & SHA2-Chain & $2^{24}$ \\ \hline
\multirow{3}{*}{\makecell{\textbf{zkCNN}\\BLS12-381\\Hyrax}} & U & LeNet5 & $2^{18}$ & \multirow{3}{*}{\makecell{\textbf{zkGPT}\\BLS12-381\\Hyrax}} & X & GPT-2 (seq 8) & $2^{23}$ \\ \cline{2-4}\cline{6-8}
 & V & VGG11 & $2^{24}$ & & Y & GPT-2 (seq 16) & $2^{25}$ \\ \cline{2-4}\cline{6-8}
 & W & VGG16 & $2^{25}$ & & Z & GPT-2 (seq 32) & $2^{26}$ \\ \hline
\end{tabular}%
}
\vspace{-1.5em}
\end{table}

\subsub{Brickell-Gordon-McCurley-Wilson (BGMW) \cite{bgmw}} BGMW reduces the number of reductions by precomputing position-weighted multiples of each base point. For each point $P_i$ and window $j$, it precomputes $2^{jc}P_i$, encoding the window's weight directly into the point. This allows all $NW$ scalar digits to be accumulated into a single bucket set, requiring only one \textit{Bucket Reduction} per MSM (or matrix row), rather than one per window as in Pippenger, and eliminating \textit{Window Reduction} entirely. The trade-off is storing $W$ precomputed points per base point and losing parallelism across windows. For matrix MSMs, parallelism across rows compensates for this loss, and the precomputed points are reused across rows.

\subsub{Straus~\cite{straus}} The original Straus algorithm avoids \textit{Bucket Reduction} by precomputing all possible weighted combinations of points within a window, requiring $2^{Nc}$ entries. We instead consider a scalable variant that precomputes the $2^c-1$ nonzero digit multiples of each base point, requiring $N(2^c-1)$ entries. During each window, scalar digits directly select the corresponding precomputed multiples, which are accumulated \textit{without} \textit{Bucket Reduction}. The per-window results are then combined through \textit{Window Reduction}. Straus is therefore attractive for small window sizes, as its precomputation storage grows exponentially with $c$.

For $N=2^{25}$ matrix MSMs, our CPU profiling measures a one-time precomputation cost of at most 2.6\,s and 805\,MB.

\subsub{Batch Addition (BA)} For binary MSMs, all nonzero scalars are one, reducing the computation to accumulating the selected points without windowing or bucketing. This requires at most $N-1$ point additions, performed through tree-based accumulation.

\subsub{Other Algorithms}
BGMW is the fixed-base form of Yao's method \cite{yao}. The Lim-Lee comb \cite{lim_lee} uses precomputation comparable to Straus but saves only doublings while requiring non-contiguous scalar bits. Bos-Coster \cite{bos_coster} repeatedly subtracts the two largest scalars, introducing serial, data-dependent operations that map poorly to pipelined hardware.

\subsection{Characterization of MSMs}\label{sec:msm_workload_char}

We characterize MSMs across six ZKP protocols (\autoref{tab:workload_letters}), covering diverse applications and established targets of architecture research. Groth16 \cite{groth16} is a widely used circuit-based zkSNARK, while HyperPlonk \cite{hyperplonk} supports more expressive arithmetic constraints. Both have been targeted by prior ZKP ASIC accelerators: Groth16 by \cite{szkp, genza_isca26, legozk, pipezk} and HyperPlonk by \cite{zkspeed, zkphire, genza_isca26}. Spartan \cite{spartan_zkp} is a sumcheck-based zkSNARK that laid the foundation for subsequent proof systems and has been targeted by NoCap \cite{nocap}. Jolt \cite{jolt_paper} is a lookup-based zkVM for proving general-purpose program execution, while zkCNN \cite{zkcnn} and zkGPT \cite{zkgpt} target Convolutional Neural Network (CNN) and LLM inference, respectively.

We analyze their MSMs along three axes: shape; scalar characteristics (width, distribution, and sparsity); and composition (size, sequence, and structure). These characteristics appear in different combinations across protocols and within individual proofs. We formalize these in \autoref{tab:taxonomy}.

\subsub{Runtime and Execution Patterns} \autoref{fig:combined_breakdown_single} shows the prover runtime breakdown on our baseline CPU (Section~\ref{sec:methodology}). MSMs account for 23--89\% of prover runtime across the protocols, with computation distributed among commitments, openings, and other MSMs. The relative contribution of each varies across protocols and problem sizes, demonstrating that no single type of MSM dominates.

\begin{figure}[t]
    \centering
    \includegraphics[width=1\linewidth]{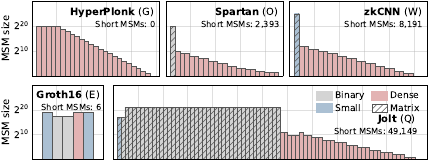}
    \vspace{-20pt}
    \caption{Size of every vector and matrix MSM in prover order for select workloads. Hatched bars are matrix MSMs. zkGPT is excluded due to its large quantity of MSMs.}
    \label{fig:fig_msm_size}
\vspace{-1.25em}
\end{figure}

\autoref{fig:fig_msm_size} plots MSM sizes in execution order, revealing distinct patterns across protocols. Groth16 performs five main MSMs, one dense and four sparse. Spartan begins with a binary matrix commitment followed by its opening step. Jolt begins with two small-scalar matrix MSMs, followed by many binary matrix MSMs with \textit{one-hot} scalars (94\% zeros). Spartan, Jolt, zkCNN, and zkGPT also issue large numbers of short MSMs containing only one to five scalar-point pairs. In Spartan, these MSMs execute sequentially, interleaved with Fiat-Shamir challenges, creating a cryptographic dependency. Consequently, they account for up to 57\% of prover runtime. In all protocols except Groth16, the opening step consists of a sequence of dense vector MSMs whose sizes halve each round.

\subsub{Shape} The shape of an MSM is largely determined by the protocol's PCS. HyperPlonk uses KZG, which produces vector MSMs over $N$ scalar-point pairs. Groth16, although it does not use a PCS, also produces vector MSMs. In contrast, Spartan, Jolt, zkCNN, and zkGPT use Hyrax or Dory, which arrange $N$ scalars into $M_r$ rows and $M_c$ columns (\autoref{fig:msm_shape}). Each row forms an independent MSM over a shared point vector, exposing row-level parallelism and point reuse. Commitment MSMs use fixed base points generated during a one-time setup. Matrix-based PCSs trade a larger commitment size ($O(\sqrt{N})$ versus KZG's single commitment) for these advantages.

\begin{figure}[t]
    \centering
    \includegraphics[width=1\linewidth]{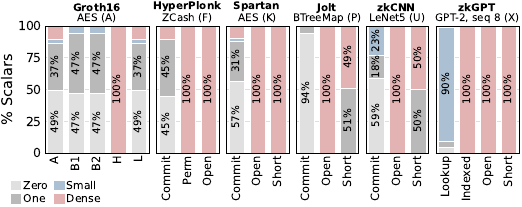}
    \vspace{-20pt}
    \caption{Distribution of scalar types in each MSM group. Witness commitments are dominated by zero and one scalars, while openings are fully dense.}
    \label{fig:msm_scalar_dist}
\vspace{-1.25em}
\end{figure}

\subsub{Scalar Characteristics}\label{sec:msm_width_sparsity_dist} \autoref{fig:msm_scalar_dist} shows the scalar distributions across different MSM groups. Commitment MSMs largely consist of binary or small scalars, while opening MSMs consistently operate on dense scalars. Many commitments contain mostly zeros and ones, whereas others, such as zkGPT's lookup commitments, contain predominantly small scalars with little sparsity. Short MSMs are generally dense. Small scalars are roughly 8--64 bits wide and can be negative. These characteristics vary across protocols and between different steps of a single proof.

\begin{figure}[t]
    \centering
    \includegraphics[width=1\linewidth]{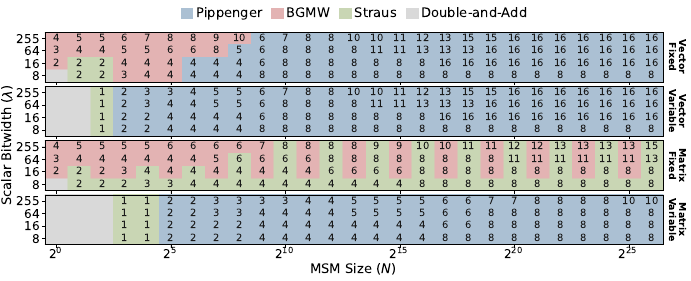}
    \vspace{-20pt}
    \caption{Lowest-cost MSM algorithm and window size across MSM size, shape, scalar width, and base-point type, using the cost model in Table~\ref{tab:msm_algo_cost}. Numbers indicate the selected window size $c \leq 16$.}
    \label{fig:best_heatmap}
\vspace{0.3em}
{\footnotesize
\centering
\captionof{table}{MSM Algorithm Cost Model}
\vspace{-7pt}
\label{tab:msm_algo_cost}
\setlength{\tabcolsep}{0.5mm}
\begin{tabular}{|l|c|c|c||c|}
\hline
 & \multicolumn{3}{c||}{\textbf{PADD Operations}} & \textbf{Stored} \\ \cline{2-4}
 & \textbf{Accumulate} & \textbf{Bucket Red.} & \textbf{Window Red.} & \textbf{Points} \\ \hline \hline
Pippenger & $NW$ & $2M_rW(2^c-1)$ & $M_r(W-1)(1+\frac{1}{2}c)$ & $M_c$ \\ \hline
BGMW & $NW$ & $2M_r(2^c-1)$ & $0$ & $WM_c$ \\ \hline
Straus & $NW$ & $0$ & $M_r(W-1)(1+\frac{1}{2}c)$ & $(2^c-1)M_c$ \\ \hline
D\&A & $\frac{3}{2}N(\lambda-1)+N$ & $0$ & $0$ & $M_c$ \\ \hline
\end{tabular}
\vspace{-2em}
}

\end{figure}
\begin{figure*}
    \centering
    \includegraphics[width=1\linewidth]{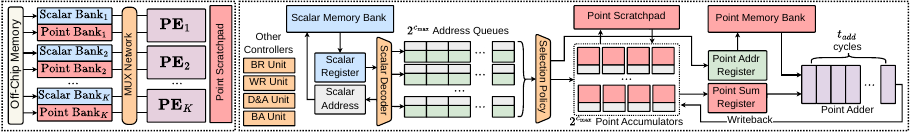}
    \vspace{-20pt}
    \caption{Tailor Architecture (Left). PE Accumulation Pipeline (Right). Scalars read from the scalar banks are decoded and pushed into the address queues. A point address and its point accumulator are then popped, and the point is read and pushed into the pipelined PADD. Each completed PADD writes back to its point accumulator.}
    \label{fig:arch}
\vspace{-1em}
\end{figure*}
\subsection{Analysis of MSM Algorithms}\label{sec:msm_algo_char}
To understand how MSM characteristics affect algorithm and configuration selection, we model the cost of each MSM algorithm from Section~\ref{sec:msm} as a function of MSM size $N$, shape ($M_r \times M_c$), and scalar width $\lambda$ (Table~\ref{tab:msm_algo_cost}). We define MSM cost as:
$$
\text{Cost} = \beta \times \text{PADD Operations} + (1 - \beta)\times \text{Stored Points}
$$
This cost model accounts for both compute and memory, weighted equally with $\beta=0.5$. The model is theoretical and hardware-agnostic: PADD counts are worst case, and a doubling counts as half an addition. Pippenger, BGMW, and Straus perform the same $NW$ accumulation additions; they differ in their reductions and precomputation requirements. Fixed base points can be precomputed once offline, while variable base points are generated during proof runtime, so precomputation does not apply to them.

\autoref{fig:best_heatmap} shows the lowest-cost algorithm and window size across MSM characteristics. \textit{Size:} Small MSMs favor Double-and-Add, Straus, and BGMW, which reduce or avoid Pippenger's repeated reductions, while large vector MSMs favor Pippenger as its reduction overhead is amortized over $N$. \textit{Shape:} Matrix MSMs reverse this trend. Pippenger repeats its reductions for every row, while BGMW and Straus reuse precomputed points across rows, making them preferable for fixed-base matrices under our model. \textit{Bitwidth:} Wider scalars delay the Pippenger crossover from $N=2^6$ (8-bit) to $N=2^9$ (255-bit); the optimal window size also grows with $N$, up to 16 bits but never exceeding $\lambda$. Finally, variable base points restrict the choice to Double-and-Add, Straus with $c=1$, and Pippenger; without precomputation, Pippenger wins all but the smallest MSMs.

Together with the heterogeneity observed in Section~\ref{sec:msm_workload_char}, these results motivate adapting algorithm and window-size selection to each MSM, while exploiting its effective scalar width and shape.

\subsection{Challenges}
Supporting heterogeneous MSMs efficiently on shared hardware presents three architectural challenges. (1) \textit{Unifying algorithms and configurations:} Pippenger, BGMW, and Straus differ in how they organize accumulations, reductions, and precomputed points. Window size and scalar width further change their resource requirements, making separate implementations costly in compute and storage. (2) \textit{Balancing resources and point reuse:} Precomputation trades reduction overhead for point storage, while matrix MSMs offer opportunities to reuse points across rows. Efficient execution requires balancing on-chip storage, accumulator allocation, and data movement to exploit these trade-offs. (3) \textit{Maintaining utilization:} Small windows, short MSMs, and dependent reductions can leave deeply pipelined PADDs underutilized. Sustaining utilization requires exposing and interleaving independent work across buckets, windows, and rows.

\section{Tailor Architecture}

To address these challenges, we introduce \textit{Tailor}, a programmable MSM architecture built around a unified accumulation datapath and a configurable multi-PE dataflow. Our key insight is that heterogeneous MSMs expose independent accumulations across buckets, windows, and matrix rows, which can be mapped onto shared hardware by dynamically partitioning a pool of point accumulators. Tailor uses configurable scalar decoding and accumulator partitioning to support different accumulation patterns and resource requirements. Across the PE array, Tailor adapts its dataflow to the different forms of parallelism exposed by each MSM shape, exploiting independent windows or matrix rows while coordinating point reuse and reduction execution.

\autoref{fig:arch} illustrates Tailor's architecture, comprising an array of PEs, a banked memory system, and a shared accumulation pipeline within each PE. We first describe the hardware organization, then explain how Tailor configures and maps heterogeneous MSMs onto these resources.

\begin{table}[b]
\small
\centering
\caption{Tailor Architecture Taxonomy.}
\vspace{-7pt}
\label{tab:arch_tax}
\setlength{\tabcolsep}{2mm}{
\begin{tabular}{|c|c|}
\hline
\textbf{Variable} & \textbf{Description} \\ \hline \hline
\text{$N$} &  Problem Size \\ \hline
\text{$\lambda_\text{max}$, $\lambda_\text{eff}$} &  Maximum and Effective scalar bitwidth \\ \hline
\text{$c_\text{max}$, $c_\text{eff}$} & Maximum and Effective Window size \\ \hline
\text{$W_\text{max}$, $W_\text{eff}$} & Maximum and Effective Windows per Scalar ($\lambda/c$) \\ \hline
\text{$K$} & Number of PEs \\ \hline
\text{$D$} &  Points/Scalars per PE \\ \hline
\text{$t_\text{add}$} & Point Addition Cycles  \\ \hline
\text{$t_\text{dbl}$} & Point Doubling Cycles \\ \hline
\text{$M_r \times M_c$} &  Rows $\times$ Columns in a Matrix \\ \hline
\text{$T_r \times T_c$} &  Rows $\times$ Columns in a Tile \\ \hline
\text{$P_r \times P_c$} &  $T_r/K$ Rows $\times$ $T_c/K$ Columns in a PE \\ \hline
\end{tabular}
}
\end{table}

\subsection{Hardware Organization}

\subsub{Accumulation Pipeline} Each Tailor PE contains a unified accumulation pipeline (\autoref{fig:arch}), built around a fully pipelined point adder (PADD) and a pool of $2^{c_{\max}}$ point accumulators. Each accumulator stores an intermediate point sum and has an associated address queue. A configurable \textit{scalar decoder} reads scalars from on-chip memory and maps each nonzero window digit to a destination accumulator and a point address. It places the point address in that accumulator's queue. Each cycle, a \textit{queue selection policy} selects a ready queue, where a queue is ready if it is nonempty and its accumulator has no addition in flight. The selected address fetches a point from the point bank, and the PADD adds it to the accumulator, writing the result back upon completion. If no queue is ready, the pipeline issues a bubble.

\subsub{Memory System} Tailor uses on-chip scalar and point memory banks connected to all PEs through a crossbar. Point banks store only the $(X,Y)$ coordinates of normalized points, with $Z=1$. The PADD produces general Jacobian points $(X,Y,Z)$ \cite{kuwar2026locus}, which are stored in the accumulators and shared point scratchpad. The shared point scratchpad preserves accumulator state while the PE accumulator pools are reused for other work. Scalars and points, including precomputed point multiples, are loaded from off-chip memory into their respective banks. Both scalar and point memories are double buffered to allow transfers to overlap computation.

\subsub{Reduction Execution} Each PE reuses its PADD and on-chip storage for accumulation and the bucket reduction (BR) and window reduction (WR) stages required by the selected algorithm. These stages execute sequentially without overlap. Dedicated BR and WR controllers schedule reduction operations on the shared PADD.

\subsection{Configurable MSM Execution}
Tailor configures each MSM through three parameters: the algorithm, effective window size $c_\text{eff}$, and effective scalar width $\lambda_\text{eff}$. The algorithm and window size are selected according to the MSM's characteristics, while the effective scalar width is determined by the widest scalar within the MSM. These parameters determine how the shared pipeline accumulates points, allocates on-chip storage resources, and performs reductions.

\begin{figure}[t]
    \centering
    \includegraphics[width=0.9\linewidth]{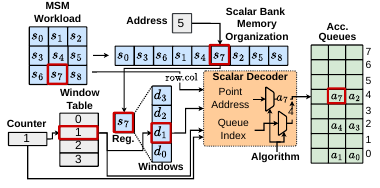}
    \vspace{-12pt}

    \caption{Scalars are read from the bank in column-major; a counter selects window digits via a table; the decoder computes algorithm-specific point addresses and queue indices.}

    \label{fig:pe_dataflow}
\vspace{-0.75em}
\vspace{1em}
{\small
\centering
\captionof{table}{Scalar Decoder Formulas}
\vspace{-7pt}
\label{tab:decoder_index}
\setlength{\tabcolsep}{1.5mm}
\begin{tabular}{|l|c|c|}
\hline
 & \textbf{Queue Index} & \textbf{Point Address} \\ \hline \hline
Pippenger & $\text{counter} \cdot 2^{c_\text{eff}} + d - 1$ & $\text{col}$ \\ \hline
BGMW & $\text{row} \cdot 2^{c_\text{eff}} + d - 1$ & $\text{col} \cdot W_\text{eff} + w$ \\ \hline
Straus & $\text{row} \cdot W_\text{eff} + w$ & $\text{col} \cdot (2^{c_\text{eff}} - 1) + d - 1$ \\ \hline
\end{tabular}
\vspace{-1em}
}

\vspace{-1em}
\end{figure}
\subsub{Algorithm Mapping} Tailor maps each algorithm onto the accumulation pipeline by configuring two decisions: which point address a window digit selects and which accumulator's address queue receives it. The scalar decoder implements these mappings and is the only algorithm-dependent stage within the accumulation pipeline. \autoref{fig:pe_dataflow} illustrates the decoding process, and Table~\ref{tab:decoder_index} gives the corresponding point-address and accumulator-index formulas.

Pippenger assigns a bucket set to each window, with each bucket occupying one point accumulator. The decoder sorts points into buckets according to their window digits by placing each base point address into the corresponding queue. Bucket Reduction combines each window's buckets, and Window Reduction combines the resulting window sums.

BGMW assigns one bucket set to each matrix row, with all windows contributing to that set. The decoder uses the window index to address a precomputed position-weighted point and the digit to select its bucket. Straus instead assigns one accumulator to each window of a row, accumulating the precomputed point multiples selected by the decoder. BGMW only requires Bucket Reduction, while Straus only requires Window Reduction.

\subsub{Window Size} The effective window size $c_\text{eff}$ determines the number of windows per scalar $W_\text{eff}=\lceil\lambda_\text{eff}/c_\text{eff}\rceil$ and the accumulator requirements of each algorithm. For Pippenger and BGMW, each bucket set requires $2^{c_\text{eff}}-1$ accumulators. Smaller windows therefore allow the fixed pool of $2^{c_\text{max}}$ accumulators to accommodate multiple windows or matrix rows, respectively. Smaller windows also shorten Bucket Reduction, at the cost of processing more windows. Straus uses one accumulator per window of each row, so the number of rows that fit in the pool depends on $W_\text{eff}$.

Tailor exploits this flexibility by repartitioning its fixed accumulator pool rather than changing the hardware provisioned for each configuration. Unlike GenZA \cite{genza_isca26}, which can require additional point transfers when buckets exceed on-chip capacity, Tailor keeps the active bucket sets on-chip, avoiding these transfers while using the remaining accumulator capacity for independent computations.

\subsub{Scalar Width} A smaller $\lambda_\text{eff}$ reduces $W_\text{eff}$, decreasing the computation performed during accumulation and reduction. Since scalar memory words are $\lambda_\text{max}$ bits wide, Tailor packs up to $\lfloor\lambda_\text{max}/\lambda_\text{eff}\rfloor$ scalars per word. A lane selector in the scalar fetch path supplies individual scalars to the decoder. This packing increases the number of scalars resident on-chip, enabling more matrix rows to be processed together.

\subsub{Programming Tailor}\label{sec:programming} Tailor selects its algorithm and window size using an offline mapping table populated through cycle-accurate simulation, indexed by MSM characteristics (size, shape, scalar width, and sparsity). This search is performed \textit{once} per hardware design. Although the cost model in Section~\ref{sec:msm_algo_char} identifies favorable configurations, it does not account for hardware attributes such as parallelism and memory transfer constraints. At runtime, the host issues an instruction specifying each MSM's size, shape, and selected configuration.

\begin{figure}[t]
    \centering
    \includegraphics[width=1\linewidth]{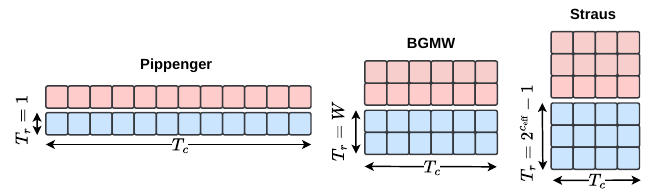}
    \vspace{-25pt}
    \caption{Tile sizes for each algorithm with $D=12$, $\lambda=4$, and $c_\text{eff}=2$. Each tile computes on the same number of scalars.}
    \label{fig:tile_sizes}
\vspace{-1.5em}
\end{figure}
\begin{figure}[t]
    \centering
    \includegraphics[width=1\linewidth]{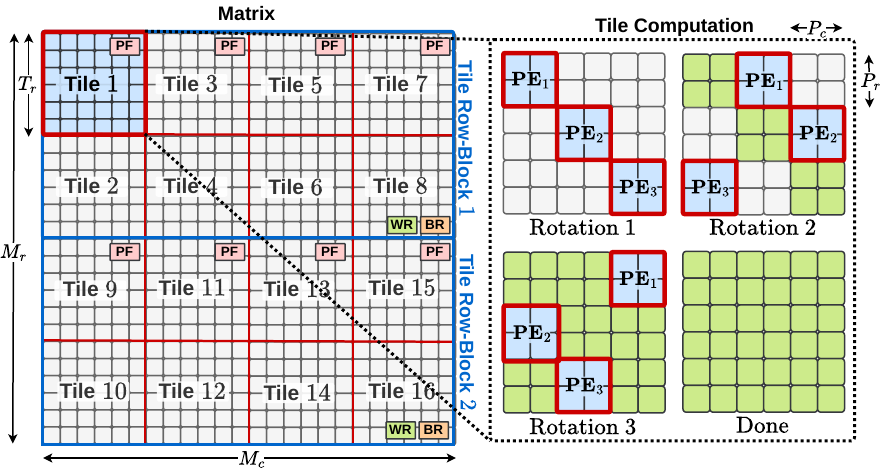}
    \vspace{-23pt}

    \caption{Matrix MSM mapping. Matrices are divided into \textit{tile row-blocks} and \textit{tiles}, processed one at a time. PEs process rows diagonally with round-robin bank rotation. Tiles fetch scalars and accumulate; some fetch points (PF) or perform window (WR) and bucket (BR) reductions. Green marks processed scalars; numbers show tile order.}

    \label{fig:matrix_par}
\vspace{-1.5em}
\end{figure}

\subsection{Multi-PE Dataflow}\label{sec:dataflow}
Tailor adapts its dataflow to distribute computation across the PE array, exploiting the different forms of parallelism available in vector and matrix MSMs. Vector mappings primarily exploit independent windows, while matrix mappings exploit independent rows that share a common point vector.

Tailor partitions MSMs into tiles processed by the $K$ PEs, with tile dimensions determined by the available on-chip storage resources. For matrix MSMs, each tile contains $T_r \times T_c$ scalars, with each PE assigned $P_r$ rows and accessing $P_c$ columns per bank rotation. Tailor uses round-robin bank rotation to coordinate scalar and point accesses across PEs, enabling concurrent execution without bank conflicts. The following sections describe how this mapping is adapted to vector and matrix MSMs.

\subsub{Matrix Mapping}\label{sec:dataflow_matrix} \autoref{fig:matrix_par} illustrates how Tailor processes a matrix MSM using tiles of $T_r \times T_c$ scalars. Each tile contains a group of rows operating on the same $T_c$ point columns. The PEs process one tile at a time, accumulating to their assigned rows.

The tile dimensions depend on the selected algorithm and its precomputation requirements (\autoref{fig:tile_sizes}). Each matrix column contains $T_r$ scalars that share the same set of points. BGMW stores $W_\text{eff}$ position-weighted points per column, while Straus stores $2^{c_\text{eff}}-1$ precomputed multiples. Precomputation changes the tile dimensions rather than requiring larger point banks. As more points are stored per column, Tailor reduces $T_c$ and increases $T_r$, allowing additional scalar rows to share the same precomputed points. This maintains a roughly equal ratio of scalars to stored points. When $\lambda_\text{eff}<\lambda_\text{max}$, scalar packing can further increase $T_r$.

\autoref{fig:matrix_par} also shows how Tailor orders tiles to reuse points. Tiles in the same column share points, so Tailor processes them downward while keeping the points loaded. Each tile covers only $T_c$ of the matrix's $M_c$ columns, so its rows remain partially accumulated until their final column tile. Tailor saves these partial sums in the point scratchpad, freeing the PE accumulators for the next tile. The rows whose partial sums fit in the scratchpad form a \textit{tile row-block}. Tailor then processes the remaining column tiles of that row-block, restoring partial sums as needed. This allows each group of points to be loaded once per row-block.

Bucket and Window Reductions contain largely sequential dependencies, limiting how frequently an individual reduction chain can issue operations to the pipelined PADD. Rather than reducing rows individually as they finish, Tailor defers reductions until the bottom-right tile of the row-block completes. This makes multiple independent reduction chains available for interleaving, allowing the PADDs to execute operations from other rows while waiting for dependent results and improving reduction efficiency.

\begin{figure}[t]
    \centering
    \includegraphics[width=1\linewidth]{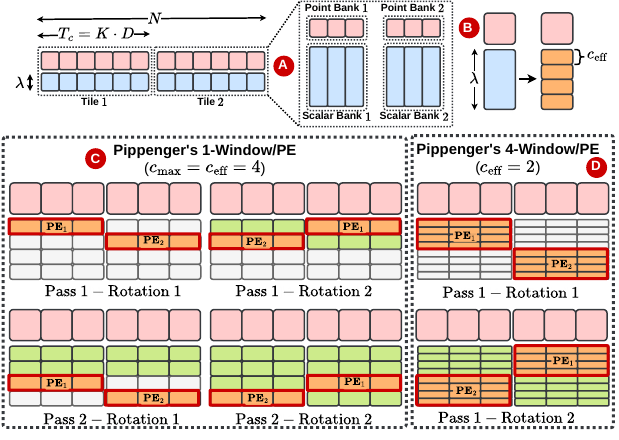}
    \vspace{-23pt}
    \caption{Vector MSM mapping. (A) Banked scalars (blue) and points (red). (B) Scalar windows (orange). (C--D) Pippenger maps one or multiple windows per PE using diagonal, round-robin bank rotation; $K<W$ requires multiple passes. Green marks processed windows.}

    \label{fig:pip_pe_rots}
\vspace{-1.5em}
\end{figure}

\subsub{Vector Mapping} \autoref{fig:pip_pe_rots} illustrates how Tailor maps vector MSMs using Pippenger's algorithm. Unlike matrix MSMs, vectors do not expose row-level parallelism. Instead, Pippenger exposes parallelism across $W_\text{eff}$ independent windows and across the buckets within each window, with all \textit{windows} sharing the same point vector. Tailor divides the vector into tiles of up to $KD$ scalar-point pairs and distributes these windows across PEs.

SZKP \cite{szkp} assigns one window to each PE at a time (\autoref{fig:pip_pe_rots}C). While this mapping exploits window-level parallelism, smaller window sizes create a challenge: each window has fewer buckets, providing fewer independent additions to the pipelined PADD. Dependencies between additions to the same bucket can therefore leave the pipeline idle. At the same time, smaller windows produce more windows per scalar, exposing additional independent work.

Tailor exploits this additional parallelism by allowing multiple windows to share a PE's accumulator pool (\autoref{fig:pip_pe_rots}D). When $c_\text{eff}=c_\text{max}$, the pool accommodates one bucket set. Reducing $c_\text{eff}$ frees accumulators that can be assigned to other windows, allowing up to $2^{c_\text{max}-c_\text{eff}}$ windows per PE. For example, \autoref{fig:pip_pe_rots}D shows four windows mapped to a PE when $c_\text{max}=4$ and $c_\text{eff}=2$. When sufficient windows are available, these windows execute concurrently within a single pass over the memory banks. Any remaining windows are processed in additional passes.

By processing multiple windows together, Tailor provides more independent accumulation operations to the shared PADD, reducing stalls caused by dependencies within individual buckets. The same principle applies during Bucket Reduction: each window has its own reduction chain, and Tailor interleaves multiple BR chains on a single PE's PADD, issuing additions from other windows while dependent results are pending.

\subsection{Additional Execution Modes}

\subsub{Batch Addition (BA)} For binary MSMs, Tailor streams the selected points into independent accumulators within each PE, allowing multiple addition chains to execute on the shared PADD. Once all points are processed, their partial sums are combined through tree-based accumulation, first within each PE and then across PEs using the shared scratchpad. Unlike prior work \cite{zkspeed, szkp}, Tailor reuses its accumulation pipeline without writing partial sums back to point memory. Batch Addition is memory-bound, as it performs approximately one point addition per loaded point.

\subsub{Double-and-Add (D\&A)} Each scalar-point multiplication executes as an independent chain in one accumulator, performing point doublings and conditional additions according to the scalar bits. These chains are distributed across PEs and interleaved within each PADD. Their results are combined through tree-based accumulation, first within each PE and then across PEs.

\subsub{Negative Scalars}\label{sec:negative_scalars_arch} Small negative scalars become full-width when represented modulo the scalar field. Tailor preserves their small width by moving the sign to the point, using $(-s)P=s(-P)$. Point negation requires only negating the $Y$ coordinate.

\subsection{End-to-End System Integration}\label{sec:e2e_system}
To evaluate end-to-end proving performance, we integrate Tailor into a full ZKP accelerator. We adopt zkPHIRE's \cite{zkphire} programmable SumCheck unit and supporting units for multilinear extension (MLE) construction and batch inversion, along with a hash unit for Fiat-Shamir challenges. A small vector unit performs the remaining element-wise modular arithmetic over the curve's scalar field. The accelerator uses a modular organization, with each proving step assigned to its corresponding hardware unit. All units share a High Bandwidth Memory (HBM) \cite{hbm3} controller.

\section{Methodology}\label{sec:methodology}
We implement and functionally verify MSM processing elements (PEs) for Tailor and zkSpeed+ using Catapult HLS 2025. Both use the same fully pipelined PADDs and modular multipliers generated with an HLS framework~\cite{kuwar2026locus}. We synthesize the RTL using Synopsys Design Compiler in GF 12nm at 1\,GHz and estimate SRAM with Synopsys 12nm Memory Compiler. Following prior work~\cite{zkspeed, zkphire, pipezk}, we primarily use the BLS12-381 curve, with PADD addition/doubling latencies of 31/16 cycles and 27/15 cycles for BN254. We use 16-entry accumulator queues with the longest-queue selection policy for high PADD utilization \cite{szkp}. We scale area and power to 7nm using factors of $1.8\times$ and $1.65\times$, respectively~\cite{zkspeed, zkphire, szkp}.

We evaluate performance using a cycle-accurate simulator modeling finite queues, backpressure, and in-flight accumulator conflicts. Off-chip memory is modeled by bandwidth, with double-buffered transfers overlapped with computation. We verify simulated results against a software reference on sample MSMs. For CPU comparisons, we use a 32-core AMD EPYC 7502 (3.35\,GHz, 296\,mm$^2$)~\cite{cpu_amd2, cpu_amd1, cpu_amd3}, and for GPU comparisons, an NVIDIA A100 (40\,GB, 1.6\,TB/s)~\cite{NVIDIA-A100-40-GPU}.

\subsub{Workloads} We collect MSM traces from the ZKP protocols in \autoref{tab:workload_letters} for characterization (Section~\ref{sec:msm_workload_char}) and evaluation (Section~\ref{sec:eval}). For Groth16 and Spartan, we use Jsnark circuits, with Groth16 proofs generated using libsnark. For Jolt, zkCNN, and zkGPT, we use examples from their codebases. Since HyperPlonk has no public compiler, we follow prior work \cite{zkspeed, zkphire, genza_isca26} and use synthetic scalars: uniform values for dense MSMs and $45\%$ zeros, $45\%$ ones, and $10\%$ dense scalars for sparse MSMs. For controlled architectural comparisons (Sections~\ref{sec:pareto} and~\ref{sec:msm_perf}), we also use synthetic MSMs: dense MSMs with 255-bit scalars drawn from $[2^{\lambda-1}, 2^{\lambda}-1]$, and MSMs with 16- or 64-bit scalars drawn from $[1, 2^{\lambda}-1]$.

\subsub{Baselines}
We compare Tailor against zkSpeed \cite{zkspeed, zkphire} and zkSpeed+, an adapted baseline that exploits row-level parallelism in matrix MSMs. zkSpeed targets a single vector MSM, making it inefficient for matrices. For matrix MSMs, zkSpeed+ divides its $K$ PEs among multiple MSM cores \cite{zkspeed}, varying the number of PEs per core (PEs/Core), with each core processing a different matrix row in parallel. For vector MSMs, it distributes the computation across its $K$ PEs and combines partial sums through tree-based accumulation. We simulate both baselines using a cycle-accurate simulator with the same PADD units, 1\,GHz clock, 16-entry per-bucket address queues with the longest-queue selection policy, and bandwidth assumptions as Tailor. Both baselines use a dedicated batch-addition datapath for binary scalars \cite{zkspeed}. We assume the same point-negation logic for negative scalars as Tailor. Both baselines retain zkSpeed's fixed-width scalar execution, while Tailor adapts its effective scalar width to each MSM.

\begin{table}[b]
\small
\centering
\caption{Design Space for Tailor and Baselines.}
\vspace{-7pt}
\label{tab:design_space}
\setlength{\tabcolsep}{1.5mm}{
\begin{tabular}{|c|c|}
\hline
\textbf{Design Knob} & \textbf{Values} \\ \hline \hline
\multicolumn{2}{|c|}{\textbf{Hardware Configuration}} \\ \hline
\textbf{PEs ($K$}) & 1, 2, 4, 8, 16, 32, 64 \\ \hline
\textbf{PEs/Core}\text{ (zkSpeed+ only)} & 1, 2, ... $K$ \\ \hline
\textbf{Maximum Window Size ($c_\text{max}$)} & 3, 4, ... 10 \\ \hline
\textbf{Points/Scalars per PE ($D$)} & 1K, 4K, 8K, 16K \\ \hline
\textbf{Bandwidth (GB/s)} & 16, 32, 64, 128, 256, 512, 1T, 2T, 4T \\ \hline \hline
\multicolumn{2}{|c|}{\textbf{Tailor Runtime Configuration}} \\ \hline
\textbf{Algorithm} & Pippenger, BGMW, Straus \\ \hline
\textbf{Effective Window Size ($c_\text{eff}$)} & 3, 4, ... $c_\text{max}$ \\ \hline
\end{tabular}
}
\end{table}

\begin{figure}[t]
    \centering
    \includegraphics[width=1\linewidth]{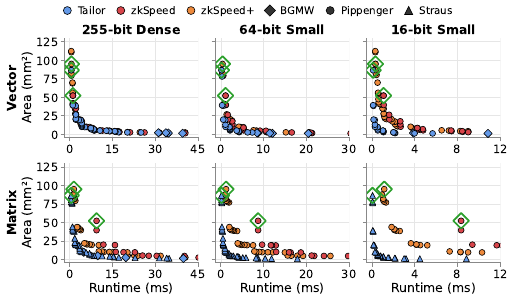}
    \vspace{-23pt}
    \caption{Pareto frontiers of Tailor, zkSpeed, and zkSpeed+ for six MSM types at $N=2^{20}$ and 1\,TB/s bandwidth. Outlines mark the selected designs.}
    \label{fig:pareto}
\vspace{-1em}
\end{figure}

\section{Evaluation}\label{sec:eval}

We explore the design space of Tailor and baseline MSM accelerators, selecting ASIC designs and comparing area and power. We then evaluate performance across MSM characteristics against ASIC, CPU, and GPU baselines, assess programmability against specialized designs, and examine compute and memory behavior. Finally, we evaluate real-protocol MSMs and end-to-end proving.

\begin{figure*}
    \centering
    \includegraphics[width=1\linewidth]{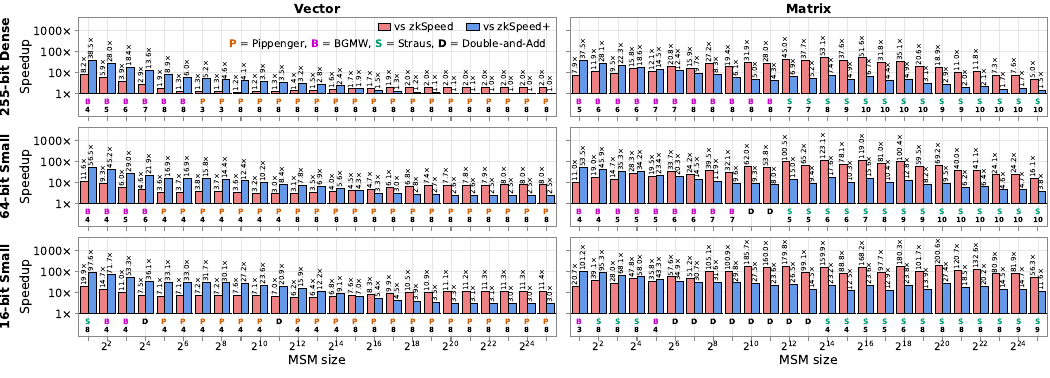}
    \vspace{-25pt}
    \caption{Tailor's speedup over zkSpeed and zkSpeed+ across MSM sizes. Both baselines run Pippenger's only; labels below each bar pair give the algorithm and window size of Tailor's fastest configuration.}
    \label{fig:per_design_curves}
\vspace{-1.25em}
\end{figure*}
\begin{figure}[t]
    \centering
    \includegraphics[width=1\linewidth]{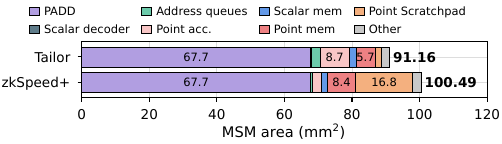}
    \vspace{-20pt}
    \caption{MSM area breakdown of the selected Tailor and zkSpeed+ designs. Areas (mm$^2$) are annotated.}
    \label{fig:area_bars}
\vspace{-2.1em}
\end{figure}
\subsection{Pareto Frontier Analysis}\label{sec:pareto}
We explore the design space of Tailor, zkSpeed, and zkSpeed+ using the parameters in \autoref{tab:design_space}. Each design is evaluated on six MSM types at $N=2^{20}$: vector and matrix MSMs with 255-bit dense, 64-bit small, and 16-bit small scalars, all with $0\%$ sparsity. Tailor selects the fastest runtime configuration for each MSM. \autoref{fig:pareto} shows the resulting Pareto frontiers at 7\,nm, using 1\,TB/s of bandwidth, consistent with prior ZKP kernel evaluations \cite{zkphire}.

On dense 255-bit vectors, all three architectures achieve similar Pareto frontiers, as Pippenger's algorithm is well suited to these MSMs. Tailor's advantages grow for matrix MSMs and smaller scalars. At comparable areas, Tailor matches or outperforms both baselines across these MSMs. Matrix MSMs benefit from algorithm selection and point reuse, while smaller scalars benefit from adapting $\lambda_\text{eff}$ and processing fewer windows. These trends are reflected in Tailor's algorithm selections. Pippenger accounts for 80\% of vector Pareto points, while Straus accounts for 91\% of matrix points. BGMW appears primarily for single-PE vector designs, where it avoids repeated Bucket Reductions, and for 255-bit matrices.

For the remaining evaluation, we select the Pareto-optimal design with the lowest geometric-mean runtime across the six MSM types for each architecture, subject to a common area budget of 127\,mm$^2$, consistent with the MSM area of prior work \cite{zkspeed, zkphire}.

\subsection{MSM Area and Power}\label{sec:area_power}
\autoref{fig:area_bars} compares the area breakdown of the selected Tailor and zkSpeed+ designs. The PADDs \cite{kuwar2026locus} are more area-efficient than those in prior work \cite{zkspeed, zkphire}, allowing both architectures to fit 64 PEs within the common area budget. Tailor occupies 91.16\,mm$^2$ and consumes 222.91\,W, compared with 100.49\,mm$^2$ and 222.96\,W for zkSpeed+. PADDs dominate both designs, accounting for 74\% and 67\% of their respective areas.

Tailor reuses its PADDs and on-chip storage across all runtime configurations, and its configurable scalar decoder occupies 0.27\% of MSM area. Overall, Tailor is 9\% smaller than zkSpeed+, primarily due to scratchpad organization: Tailor shares a single point scratchpad across all PEs, whereas zkSpeed+ replicates its scratchpad across 32 MSM cores.

\subsection{MSM Performance}\label{sec:msm_perf}
\subsub{ASIC Comparison}\label{sec:compare_asic}
\autoref{fig:per_design_curves} compares the selected Tailor, zkSpeed, and zkSpeed+ designs on vector and matrix MSMs with 255-, 64-, and 16-bit scalars, over $N=2^{1}$ to $2^{25}$, using fixed base points. We report geometric-mean speedups over $N=2^{14}$ to $2^{25}$, representative of the problem size range in \autoref{tab:workload_letters}. Tailor uses the runtime configurations selected through the programming methodology described in Section~\ref{sec:programming}. For dense 255-bit MSMs, Tailor achieves $1.9\times$/$1.25\times$ speedups on vectors and $18.5\times$/$3.1\times$ on matrices over zkSpeed\slash zkSpeed+, respectively. At smaller vector sizes, Tailor selects BGMW, which accumulates all scalar windows into a single bucket set, avoiding Pippenger's repeated Bucket Reductions. As MSM size increases, Tailor switches to Pippenger and increases the window size to better amortize reduction overhead. On dense matrices, Tailor favors BGMW at smaller sizes and Straus at larger sizes, exploiting shared precomputed points across rows while reducing Pippenger's repeated reduction overhead.

Scalar-width adaptation further improves small-scalar performance. At 16 bits, Tailor achieves geometric-mean speedups of $10.0\times$/$4.0\times$ on vectors and $114.6\times$/$17.3\times$ on matrices over zkSpeed\slash zkSpeed+. For small-scalar MSMs, the baseline architectures retain full-width execution, whereas Tailor exploits the effective scalar width to reduce the number of processed windows and increase on-chip scalar density. For short MSMs with small scalars, Tailor can also select Double-and-Add, avoiding bucketing entirely. For example, Tailor selects it for 16-bit matrix MSMs at $N=2^{6}$--$2^{13}$.

\begin{table}[b]
\small
\centering
\caption{Vector and matrix MSM runtime (ms) on CPU, GPU, and Tailor, with Tailor's speedup over the CPU and ICICLE.}
\vspace{-7pt}
\label{tab:msm_cpu_gpu}
\resizebox{\columnwidth}{!}{%
\setlength{\tabcolsep}{1mm}%
\begin{tabular}{|c|c|c|c|c||c|c|c|}
\hline
 & \multicolumn{4}{c||}{\textbf{Vector}} & \multicolumn{3}{c|}{\textbf{Matrix}} \\ \hline
\textbf{Size} & \textbf{CPU} & \shortstack{\textbf{GPU}\\\textbf{ICICLE}} & \shortstack{\textbf{GPU}\\\textbf{Prior}} & \shortstack{\textbf{Tailor}\\\textbf{(vs CPU/GPU)}} & \textbf{CPU} & \shortstack{\textbf{GPU}\\\textbf{ICICLE}} & \shortstack{\textbf{Tailor}\\\textbf{(vs CPU/GPU)}} \\ \hline \hline
$2^{18}$ & 176 & 21.49 & 20 \cite{gzkp} & 0.14 (\textbf{1218$\times$}/\textbf{149$\times$}) & 295 & 14.38 & 0.12 (\textbf{2465$\times$}/\textbf{120$\times$}) \\ \hline
$2^{19}$ & 349 & 29.98 & -- & 0.28 (\textbf{1264$\times$}/\textbf{109$\times$}) & 496 & 20.97 & 0.23 (\textbf{2176$\times$}/\textbf{92$\times$}) \\ \hline
$2^{20}$ & 673 & 46.08 & 62 \cite{gzkp} & 0.54 (\textbf{1248$\times$}/\textbf{85$\times$}) & 1000 & 36.87 & 0.50 (\textbf{2013$\times$}/\textbf{74$\times$}) \\ \hline
$2^{21}$ & 1533 & 76.58 & -- & 1.07 (\textbf{1439$\times$}/\textbf{72$\times$}) & 1805 & 67.38 & 0.98 (\textbf{1848$\times$}/\textbf{69$\times$}) \\ \hline
$2^{22}$ & 2787 & 136.82 & 58 \cite{distmsm} & 2.12 (\textbf{1316$\times$}/\textbf{65$\times$}) & 3570 & 129.98 & 1.82 (\textbf{1961$\times$}/\textbf{71$\times$}) \\ \hline
$2^{23}$ & 5646 & 250.96 & -- & 4.22 (\textbf{1337$\times$}/\textbf{59$\times$}) & 6434 & 249.25 & 3.62 (\textbf{1777$\times$}/\textbf{69$\times$}) \\ \hline
$2^{24}$ & 11363 & 493.49 & 234 \cite{distmsm} & 8.43 (\textbf{1348$\times$}/\textbf{59$\times$}) & 12792 & 518.54 & 6.92 (\textbf{1849$\times$}/\textbf{75$\times$}) \\ \hline
$2^{25}$ & 23192 & 1106.40 & -- & 16.85 (\textbf{1376$\times$}/\textbf{66$\times$}) & 23235 & 994.57 & 13.81 (\textbf{1683$\times$}/\textbf{72$\times$}) \\ \hline
\end{tabular}%
}
\end{table}

\subsub{CPU Comparison}
\autoref{tab:msm_cpu_gpu} compares Tailor against arkworks \cite{arkworks} on 255-bit dense vector and matrix MSMs over BLS12-381, using 32 threads on an AMD EPYC 7502 CPU. Tailor achieves geometric-mean speedups of $1317\times$ and $1958\times$ on vectors and matrices, respectively, with matrix speedup reaching $2465\times$ at $N=2^{18}$.

\subsub{GPU Comparison}
We compare against ICICLE \cite{icicle} on an NVIDIA A100, which implements Pippenger's algorithm. Since prior GPU work \cite{distmsm, cuzk, gzkp, elasticmsm} targets vector MSMs, we use ICICLE's support for batching matrix rows over shared points rather than processing each row independently. We use the fastest ICICLE configuration for each MSM size. Tailor achieves approximately $79\times$ geometric-mean speedup over ICICLE on both vector and matrix MSMs. Against published vector MSM results from GZKP \cite{gzkp} and DistMSM \cite{distmsm}, Tailor achieves a $59\times$ geometric-mean speedup.

\subsection{Programmability vs. Specialization}\label{sec:prog_vs_spec}
\begin{figure}[t]
    \centering
    \includegraphics[width=1\linewidth]{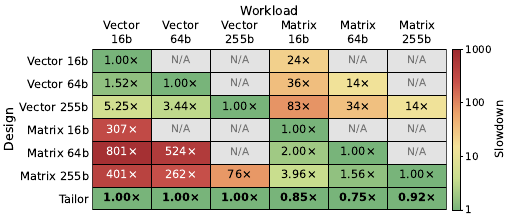}
    \vspace{-20pt}
    \caption{Geometric-mean slowdown relative to the specialized design for each MSM type over $N=2^{14}$--$2^{25}$. Unsupported combinations are shaded gray.}
    \label{fig:programmability}
    \vspace{-1em}
\end{figure}

To evaluate the value of programmability, we compare Tailor against a specialized design for each of the six MSM types. Each specialized design fixes its algorithm, window size, and scalar width at design time, selecting the fastest configuration at $N=2^{20}$ from Tailor's design space, at the same area as the selected Tailor design. \autoref{fig:programmability} reports each design's geometric-mean slowdown relative to the corresponding specialized design over $N=2^{14}$--$2^{25}$. Designs fixed to smaller scalars cannot execute MSMs with larger scalars.

\begin{figure}[t]
    \centering
    \includegraphics[width=1\linewidth]{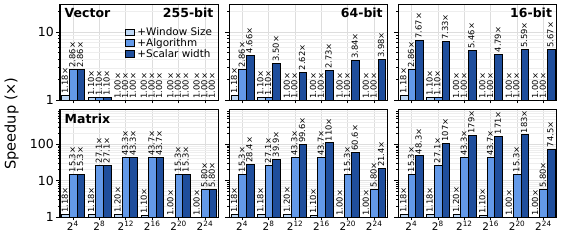}
    \vspace{-20pt}
    \caption{Cumulative speedup from enabling window-size, algorithm, and effective scalar-width selection, relative to fixed Pippenger execution ($c=8$, 255-bit scalars).}
    \label{fig:algo_window_ablation}
\vspace{-1.25em}
\end{figure}
Specialized designs perform well on their target MSMs but can lose substantial efficiency when MSM characteristics change.
The largest penalties arise from differences in MSM shape. Matrix-specialized designs are $318\times$ slower on vectors on average, as they load multiple precomputed points per input point, but a vector MSM cannot amortize these loads across multiple scalar rows. Conversely, vector-specialized designs are $28\times$ slower on matrices because they cannot exploit shared points across rows. Differences in scalar width have a smaller impact, averaging $2.6\times$ slowdown.

In contrast, a single Tailor design matches the specialized designs on vectors and outperforms them on matrices by 1.19$\times$ on average across the evaluated sizes.
This advantage arises because the optimal configuration changes with MSM size. While specialized designs retain the configuration selected at $N=2^{20}$, Tailor adapts its algorithm and window size across sizes without changing the underlying hardware.

\autoref{fig:algo_window_ablation} examines runtime configuration using Tailor fixed to Pippenger's algorithm with $c=8$ and 255-bit scalars as the reference, representative of configurations favored by large, dense vector MSMs. Window-size and algorithm selection are first evaluated with fixed 255-bit execution, followed by execution at each MSM's effective scalar width. Window-size selection alone provides up to $1.20\times$ speedup, with gains concentrated at smaller sizes. Adding algorithm selection is particularly beneficial for matrices, reaching $5.8$--$43.7\times$ cumulative speedups by selecting BGMW for smaller MSMs and Straus for larger ones. On vectors, algorithm selection benefits the smallest plotted size, reaching $2.86\times$. With effective scalar-width selection, cumulative speedups reach up to $7.67\times$ on vectors and $183\times$ on matrices at 16 bits, where Double-and-Add also becomes favorable for short MSMs. These results illustrate how runtime configuration adapts the same hardware to different MSM characteristics.

\begin{figure}[t]
    \centering
    \includegraphics[width=1\linewidth]{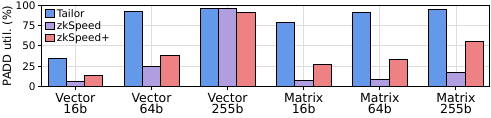}
    \vspace{-20pt}
    \caption{PADD utilization of Tailor, zkSpeed, and zkSpeed+ for six MSM types at $N=2^{20}$.}
    \label{fig:padd_util}
\vspace{-1em}
\end{figure}
\begin{figure}[t]
    \centering
    \includegraphics[width=1\linewidth]{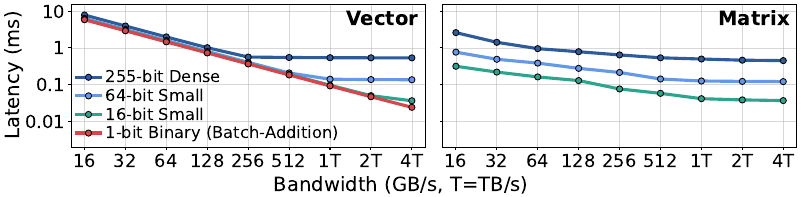}
    \vspace{-20pt}
    \caption{Tailor's sensitivity to off-chip bandwidth for vector and matrix MSMs at $N=2^{20}$, using the fastest runtime configuration for each MSM.}
    \label{fig:bandwidth}
    \vspace{-1.5em}
\end{figure}

\subsection{Compute and Memory Analysis}\label{sec:bw_sens}
\subsub{Utilization} \autoref{fig:padd_util} compares PADD utilization across the six MSM types at $N=2^{20}$, measured as occupied PADD issue slots across all PEs over the full MSM runtime, including memory waits and reductions. Tailor achieves 79--97\% utilization across all but 16-bit vectors, while zkSpeed and zkSpeed+ fall to 6--56\% outside dense vectors. These results demonstrate how Tailor's configurable accumulation and dataflow expose independent work across buckets, windows, and matrix rows, keeping the PADDs busy across diverse MSMs. For 16-bit vectors, fewer scalar operations leave less computation to overlap memory transfers, limiting overall utilization despite efficient accumulation.

\subsub{Memory Bandwidth} \autoref{fig:bandwidth} evaluates Tailor's sensitivity to off-chip bandwidth at $N=2^{20}$. Matrix MSMs are relatively insensitive to bandwidth because each loaded point is reused across multiple scalar rows, amortizing memory transfers over more computation. At 1\,TB/s, all matrix MSMs are within 13\% of their latency at 4\,TB/s. In contrast, vector MSMs lack row-level point reuse, making their bandwidth sensitivity dependent on scalar width. Wider scalars require more accumulation work, allowing more memory transfer time to be hidden by computation. Dense 255-bit vectors become predominantly compute-bound at 256\,GB/s and 64-bit vectors at 1\,TB/s, while 16-bit vectors remain memory-bound at 2\,TB/s. Binary vector MSMs remain memory-bound across the evaluated bandwidth range, requiring roughly one addition per loaded point.

\subsection{Real-World MSM Performance}\label{sec:real_workloads}
\begin{figure}[t]
    \centering
    \includegraphics[width=1\linewidth]{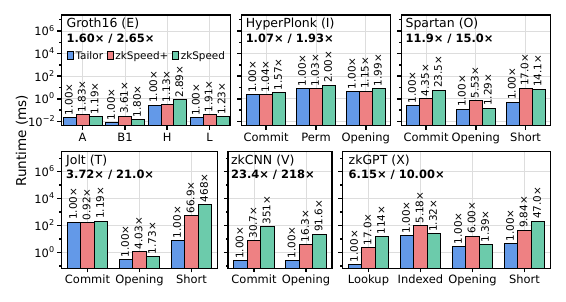}
    \vspace{-25pt}
    \caption{Per-stage MSM runtime across six ZKP protocols on Tailor, zkSpeed+, and zkSpeed. Stage and total speedups are annotated.}
    \label{fig:real_workloads}
    \vspace{-1em}
\end{figure}

We evaluate MSM sequences from selected workloads across six ZKP protocols (\autoref{tab:workload_letters}), using each protocol's native curve. We exclude Groth16's G2 MSM B2, which requires a separate, larger PADD \cite{szkp}. \autoref{fig:real_workloads} reports per-stage MSM runtimes and total speedups over zkSpeed+ and zkSpeed.

Across the selected workloads, Tailor achieves geometric-mean speedups of $4.7\times$ and $12.3\times$ over zkSpeed+ and zkSpeed, respectively. Matrix-based protocols benefit most, reaching $23.4\times$ over zkSpeed+ on zkCNN. Tailor uses Straus with reduced scalar widths for small-scalar matrix commitments in zkCNN and zkGPT, while BGMW accelerates Spartan's short MSMs, yielding an $11.9\times$ overall speedup. Jolt's short MSMs benefit from Double-and-Add, but its one-hot commitments use batch addition across all designs, limiting its speedup to $3.72\times$. Groth16 and HyperPlonk achieve smaller gains of $1.60\times$ and $1.07\times$, respectively, as their large, dense vector MSMs already favor Pippenger. Tailor benefits these protocols by selecting smaller windows and BGMW for short MSMs.

\subsection{End-to-End Prover Evaluation}\label{sec:e2e}
\begin{table}[b]
\vspace{-1em}
\small
\caption{Area and power of the End-to-End ZKP System.}
\vspace{-7pt}
\label{tab:e2e_area}
\centering
\setlength{\tabcolsep}{1mm}
\begin{tabular}{lrr}
\hline
& \textbf{Area (mm²)} & \textbf{Average Power (W)} \\ \hline
Tailor MSM (64 PEs) & 72.11 & 220.31 \\
SumCheck (16 PEs) & 5.53 & 15.15 \\
Multifunction Forest (80 Trees) & 15.81 & 43.29 \\
Vector Unit & 1.19 & 3.25 \\
Other & 6.13 & 6.87 \\
\textbf{Total Compute} & \textbf{100.77} & \textbf{288.87} \\ \hline
SRAM (57.83 MB) & 27.63 & 3.77 \\
Interconnect & 26.42 & 14.83 \\
HBM3 (2 PHYs) & 59.20 & 63.60 \\
\textbf{Total Memory System} & \textbf{113.25} & \textbf{82.20} \\ \hline
\textbf{Total} & \textbf{214.02} & \textbf{371.07} \\ \hline
\end{tabular}
\end{table}

\begin{table}[t]
\small
\centering
\caption{End-to-end prover runtime (ms) on CPU and Tailor, with Tailor's speedup over the CPU.}
\vspace{-7pt}
\label{tab:e2e_cpu}
\setlength{\tabcolsep}{0.8mm}
\resizebox{\columnwidth}{!}{%
\begin{tabular}{|l|c|c|c||l|c|c|c|}
\hline
\multicolumn{4}{|c||}{\textbf{HyperPlonk}} & \multicolumn{4}{c|}{\textbf{Spartan}} \\ \hline
\textbf{Workload} & \textbf{Size} & \textbf{CPU} & \textbf{Tailor} & \textbf{Workload} & \textbf{Size} & \textbf{CPU} & \textbf{Tailor} \\ \hline \hline
Zcash & $2^{17}$ & 1429 & 0.72 (\textbf{1973$\times$}) & AES & $2^{14}$ & 555 & 0.59 (\textbf{945$\times$}) \\ \hline
Auction & $2^{20}$ & 8619 & 4.82 (\textbf{1788$\times$}) & SHA2 & $2^{15}$ & 660 & 0.64 (\textbf{1036$\times$}) \\ \hline
Zexe's Rec. Ckt & $2^{22}$ & 37469 & 18.76 (\textbf{1997$\times$}) & RSA & $2^{17}$ & 1277 & 1.03 (\textbf{1242$\times$}) \\ \hline
Rollup 10 Tx & $2^{23}$ & 74052 & 37.33 (\textbf{1984$\times$}) & Merkle & $2^{17}$ & 1200 & 0.91 (\textbf{1313$\times$}) \\ \hline
Rollup 25 Tx & $2^{24}$ & 145535 & 74.46 (\textbf{1955$\times$}) & Auction & $2^{20}$ & 4702 & 2.08 (\textbf{2258$\times$}) \\ \hline
\end{tabular}%
}
\vspace{-1.25em}
\end{table}

\begin{table}[t]
\small
\centering
\caption{End-to-end prover runtime (ms) with zkSpeed+ and Tailor MSM units, with Tailor's speedup over zkSpeed+.}
\vspace{-7pt}
\label{tab:e2e_asic}
\setlength{\tabcolsep}{0.8mm}
\resizebox{\columnwidth}{!}{%
\begin{tabular}{|l|c|c|c||l|c|c|c|}
\hline
\multicolumn{4}{|c||}{\textbf{HyperPlonk}} & \multicolumn{4}{c|}{\textbf{Spartan}} \\ \hline
\textbf{Workload} & \textbf{Size} & \textbf{zkSpeed+} & \textbf{Tailor} & \textbf{Workload} & \textbf{Size} & \textbf{zkSpeed+} & \textbf{Tailor} \\ \hline \hline
Zcash & $2^{17}$ & 1.23 & 0.72 (\textbf{1.70$\times$}) & AES & $2^{14}$ & 7.72 & 0.59 (\textbf{13.16$\times$}) \\ \hline
Auction & $2^{20}$ & 5.44 & 4.82 (\textbf{1.13$\times$}) & SHA2 & $2^{15}$ & 8.23 & 0.64 (\textbf{12.91$\times$}) \\ \hline
Zexe's Rec. Ckt & $2^{22}$ & 19.58 & 18.76 (\textbf{1.04$\times$}) & RSA & $2^{17}$ & 9.75 & 1.03 (\textbf{9.48$\times$}) \\ \hline
Rollup 10 Tx & $2^{23}$ & 38.37 & 37.33 (\textbf{1.03$\times$}) & Merkle & $2^{17}$ & 9.81 & 0.91 (\textbf{10.74$\times$}) \\ \hline
Rollup 25 Tx & $2^{24}$ & 75.93 & 74.46 (\textbf{1.02$\times$}) & Auction & $2^{20}$ & 12.99 & 2.08 (\textbf{6.24$\times$}) \\ \hline
\end{tabular}%
}
\vspace{-1.5em}
\end{table}

We evaluate HyperPlonk and Spartan as representative protocols with contrasting MSM characteristics: HyperPlonk primarily uses large, dense vector MSMs, while Spartan combines matrix commitments with numerous short MSMs. We use the full prover accelerator (Section~\ref{sec:e2e_system}) with 2\,TB/s of off-chip bandwidth, comparing systems that differ only in their MSM units (Tailor or zkSpeed+). With Tailor, the system occupies 214.02\,mm$^2$ and consumes 371.07\,W (\autoref{tab:e2e_area}), compared with 223.35\,mm$^2$ and 371.12\,W with zkSpeed+.

\autoref{tab:e2e_cpu} compares against the CPU, where Tailor achieves geometric-mean proving speedups of $1938\times$ on HyperPlonk and $1292\times$ on Spartan. Against the ASIC system with zkSpeed+ (\autoref{tab:e2e_asic}), Tailor achieves a $10.2\times$ geometric-mean speedup on Spartan. This gain is driven primarily by Spartan's numerous sequential short MSMs, where BGMW avoids Pippenger's repeated Bucket Reductions. In contrast, HyperPlonk achieves a smaller $1.16\times$ geometric-mean speedup because its large, dense vector MSMs already favor Pippenger. The speedup reaches $1.70\times$ on the smallest workload, where short opening MSMs account for a larger fraction of runtime.
These results show that Tailor's per-MSM programmability can translate into substantial end-to-end gains, particularly for protocols containing heterogeneous MSMs.

\section{Related Works}
Prior work on MSM acceleration has primarily focused on Pippenger's algorithm, in ASIC \cite{szkp, pipezk, priormsm, zkspeed, zkphire, gypsophila, myosotis, msmac, legozk} and GPU \cite{gzkp, cuzk, distmsm}.
These designs primarily target vector MSMs, where extending execution to matrices by processing rows independently incurs repeated reductions.
GenZA \cite{genza_isca26} also selects the window size per MSM, but raises it for large MSMs at the cost of memory traffic, whereas Tailor lowers it for small MSMs. GenZA and LegoZK \cite{legozk} also support variable bitwidths, but for the field of different curves. Tailor differs from prior MSM accelerators by unifying multiple MSM algorithms on a shared accumulation datapath, while adapting its resource partitioning and dataflow to MSM shape and effective scalar width.

Beyond MSM, several accelerators target the full prover: PipeZK \cite{pipezk}, SZKP \cite{szkp}, and LegoZK \cite{legozk} for Groth16, zkSpeed \cite{zkspeed} and zkPHIRE \cite{zkphire} for HyperPlonk, and GenZA \cite{genza_isca26} for multiple protocols. Other accelerators target hash-based commitment schemes \cite{nocap, unizk, mtu, batchzk}, which avoid elliptic-curve arithmetic and prove faster but produce larger proofs; Tailor targets the ECC-based class.

\section{Conclusion}
We presented \textit{Tailor}, a programmable MSM architecture that supports diverse algorithms and runtime configurations on a shared accumulation datapath. Through flexible resource allocation and matrix-aware execution that exploits point reuse and parallelism, Tailor efficiently supports heterogeneous MSMs while maintaining strong performance on dense vectors. Across real ZKP protocols, it achieves a $4.7\times$ geometric-mean speedup, translating into a $10.2\times$ end-to-end proving speedup on Spartan. These results demonstrate that adapting how shared hardware executes each MSM enables efficient acceleration across diverse ZKP protocols.

\begin{acks}
OpenAI’s ChatGPT and Codex, and Anthropic’s Claude, assisted with cycle-accurate simulator implementation from author defined specifications, experimental data processing and visualization, and manuscript editing. The authors determined the experimental meth\-odology, architectural decisions, and evaluation objectives.
\end{acks}

\balance
\bibliographystyle{ACM-Reference-Format}
\bibliography{refs}

\end{document}